\documentclass[conference]{IEEEtran}
\usepackage{blindtext, graphicx, tabularx, booktabs, amsmath, siunitx, subfigure, xcolor}

\usepackage{cite}

\usepackage{color}
\usepackage{soul}
\usepackage{flushend}

\newcolumntype{R}[1]{>{\raggedright\arraybackslash}p{#1}}

\IEEEoverridecommandlockouts

\begin{document}
%
\title{

A Terabit Hybrid FPGA-ASIC Platform for Switch Virtualization


\thanks{

This work was financed in part by the Coordena\c{c}\~{a}o de Aperfei\c{c}oamento de Pessoal de N\'{i}vel Superior - Brasil (CAPES) - Finance Code 001, CNPq, and FAPERGS. This work also received funding from Grant \#2020/05183-0, São Paulo Research Foundation (FAPESP). This work has also been partially conducted in the project ``ICT programme'' which was supported by the European Union through the European Social Fund. It was also partially supported by the Estonian Research Council grant MOBERC35.}

}

 \author{
     \IEEEauthorblockN{
         Mateus Saquetti\IEEEauthorrefmark{1},
         Raphael M. Brum\IEEEauthorrefmark{1},
         Bruno Zatt\IEEEauthorrefmark{2},
         Samuel Pagliarini\IEEEauthorrefmark{3},
         Weverton Cordeiro\IEEEauthorrefmark{1},
         Jose R. Azambuja\IEEEauthorrefmark{1}
     }
  
      \IEEEauthorblockA{
          \IEEEauthorrefmark{1}Federal University of Rio Grande do Sul (UFRGS) - Institute of Informatics - School of Engineering
          \\\{mateus.tirone, brum, weverton.cordeiro, azambuja\}@ufrgs.br
      }
    
      \IEEEauthorblockA{
          \IEEEauthorrefmark{2}Federal University of Pelotas (UFPEL) - Center for Technological Development
          \\zatt@inf.ufpel.edu.br
      }
    
      \IEEEauthorblockA{
          \IEEEauthorrefmark{3}Tallinn University of Technology (TalTech) - Department of Computer Systems
          \\samuel.pagliarini@taltech.ee
      }
}

\maketitle

\begin{abstract}

The roll-out of technologies like 5G and the need for multi-terabit bandwidth in backbone networks requires networking companies to make significant investments to keep up with growing service demands. For lower capital expenditure and faster time-to-market, companies can resort to anything-as-a-service providers to lease virtual resources. Nevertheless, existing virtualization technologies are still lagging behind next-generation networks' requirements. This paper breaks the terabit barrier by introducing a hybrid FPGA-ASIC architecture to virtualize programmable forwarding planes. In contrast to existing solutions, our architecture involves an ASIC that multiplexes network flows between programmable virtual switches running in an FPGA capable of full and partial reconfiguration, enabling virtual switch hot-swapping. Our evaluation shows the feasibility of a switch virtualization architecture capable of achieving a combined throughput of 3.2 Tbps by having up to 26 virtual switch instances in parallel with low resource occupation overhead.

\end{abstract}
    

\section{Introduction}
\label{sec_introduction}


The networking community has long taken advantage of virtualization to realize a multitude of services dubbed XaaS (\emph{anything-as-a-service})~\cite{bari2012data,jain2013network}. Popular examples include \emph{infrastructure}, \emph{platform}, and \emph{software} as a service (IaaS, PaaS, and SaaS, respectively). In this context, virtualization has been used to leverage abstractions of several networking components (e.g. links~\cite{bari2012data,van2017defining}), forwarding elements (e.g., switches)~\cite{openvswitch-nsdi15,pisces-sigcomm-16,picnic-sigcomm2019}, servers~\cite{multi-tenant-networks-nsdi-2014}, management units~\cite{covisor-nsdi2015,blenk2015survey,afolabi2018network}, etc.

With the emergence of network programmability~\cite{rmt-sigcomm-2013,Bosshart:2014:P4,lyra-sigcomm2020} and the possibility to redefine the behavior of forwarding elements through home-brewed software~\cite{Bosshart:2014:P4,Song:2013:PFU:2491185.2491190,Lyra,npl}, researchers have been investigating how to leverage abstractions of virtual programmable forwarding elements~\cite{hancock2016hyper4,hypervdp,p4visor,reprogrammableswitches2019,virtp4_sigcomm19,p4vbox}. Their motivations for programmable virtual forwarding elements are manifold. For example, with the roll-out of 5G and the need for multi-terabit bandwidth in backbone networks~\cite{abdelwahab2016network,condoluci2018softwarization}, telecommunication companies interested in providing 5G coverage to its users may avoid significant capital expenditure with the installation of radio-base stations and equipment purchase by leasing them from an infrastructure provider~\cite{reprogrammableswitches2019}. In this case, the infrastructure provider may create virtual instances of programmable forwarding elements to its customers. Each customer can then redefine the virtual forwarding element's behavior, using one's set of networking protocols to enable a seamless integration into the customer's network.

To realize the full potential of virtualization for programmable forwarding planes and establish a novel class of XaaS called \emph{programmable switches-as-a-service}~\cite{reprogrammableswitches2019}, the research community has been addressing several challenges in the path of a fully-fledged virtualization solution. In prior investigations, researchers have devised solutions to (i) accommodate various instances of virtual switches within a single physical programmable switch hardware, either using switch emulation or switch program composition~\cite{hancock2016hyper4,hypervdp,p4visor}, (ii) enable virtual switch isolation and hot-swapping of virtual switch instances~\cite{virtp4_sigcomm19}, and (iii) provide proper abstractions of management channels for tenant-independent switch operation~\cite{vifc_sigcomm20}.

Despite the progress achieved, existing solutions failed to deliver virtual programmable switches with similar performances to bare metal ones. For example, solutions based on emulation of virtual switches or switch program composition require additional forwarding tables to implement flow steering between virtual switch instances, increasing memory overhead, and imposing severe penalties to packet latency and virtual switch throughput~\cite{p4vbox}. On the other hand, solutions capable of providing virtual switch isolation have not approached an architectural design capable of achieving line-rate speeds while ensuring virtual switch programmability.

In this paper, we bridge this gap by proposing a hybrid FPGA-ASIC platform for switch virtualization. In contrast to existing work, we take advantage of increased clock frequencies with an ASIC and combine it with the advantages of reconfigurability and flexibility provided by an FPGA -- the integrated platform delivers virtual programmable switches that can work at line rate up breaking the terabit barrier. To the best of our knowledge, no virtualization architecture has proven capable of delivering such throughput while delivering proper abstractions of virtual switches with true tenant isolation, providing virtual switch hot-swapping, and supporting backend scalability~\cite{SUME_micro,TOMAHAWK,TOFINO,SILICON_ONE}.

The remainder of this paper is organized as follows. Section \ref{sec_related} presents related works. In Section~\ref{sec_virtualization}, we discuss our reference architecture for programmable switch virtualization. In Section~\ref{sec_architecture}, we describe our implementation of a hybrid FPGA-ASIC platform and emphasize the core technical aspects that enable it to break the terabit mark for virtual switches. In Section~\ref{sec_results}, we present the results achieved with our prototypical implementation, highlighting area occupation and throughput. Finally, we close the paper in Section~\ref{sec_conclusion} with concluding remarks and directions for future research.

\section{Related Works}
\label{sec_related}


    Even though the concept of switch virtualization is well-studied, the recent development of Domain-Specific Languages (DSL) targeting programmable forwarding planes has brought a renewed interest in the area, especially when targeting the data-plane virtualization. The first work in the literature to provide a forwarding plane virtualization platform was P4VBox~\cite{p4vbox}. The authors realized an open-source platform based on the canonical NetFPGA reference design~\cite{Ibanez:2019:PWL:3289602.3293924} and adapted it for multiple switch instances. Their approach was able to instantiate up to 13 virtual switches and achieve a 40 Gbps throughput. Following this approach, MTPSA was proposed under the same premises: to provide multi-tenant portable switch architecture over the NetFPGA plataform~\cite{Noa}. Still, MTPSA is restricted to a 40 Gbps throughput due to its FPGA implementation.
    
    Aiming to improve FPGA performance and improve previous works' throughputs, the literature draws the technology enabler to boost the NetFPGA platform to a 100 Gbps throughput with the NetFPGA SUME board based on a Xilinx Virtex-7 690T FPGA device~\cite{SUME_micro}. To do so, the authors propose to use a set of the 30 serial links running at up to 13.1 Gbps. Still, the combined throughput falls short of 0.4 Tbps. Finally, newer FPGA devices such as the Xilinx UltraScale+ could improve performance and boost resource availability, but an improvement in an order of magnitude would be required to break the terabit wall.
    
    When considering ASIC platforms for programmable data-planes, virtualization can only be achieved by replicating the complete switch hardware. In other words, one must design a DSL-enabled ASIC switch and replicate it for each additional virtualized switch. Thus, for achieving multiple virtual switch instances, one must predefine the maximum number of virtual switches and design the ASIC accordingly. This approach has been adopted by several industry players, such as Broadcom Tomahawk~\cite{TOMAHAWK}, Intel Tofino~\cite{TOFINO}, and Cisco Silicon One~\cite{SILICON_ONE}. Even though these approaches reach the terabit throughput, their virtualization proposal is limited to a few instances, and resource waste is high, as one must design the switch hardware according to the worst-case switch instance.
    
    This paper proposes a hybrid FPGA-ASIC platform to combine the best of ASIC and FPGA technologies while minimizing their downsides. To do so, we take advantage of the FPGA flexibility for implementing multiple parallel virtual switches and offload the remaining static hardware blocks to an ASIC platform. Thus, we can forward packets in multiple parallel virtual switches at gigabit throughputs in the FPGA-side and serialize their combined terabit throughput through the ASIC-side. By doing so, we can achieve up to 3.2 Tbps while running up to 26 virtual switches.

\section{Proposed Platform Architecture}
\label{sec_virtualization}

    \begin{figure}[t!]
    \centering
    \includegraphics[width=1\linewidth]{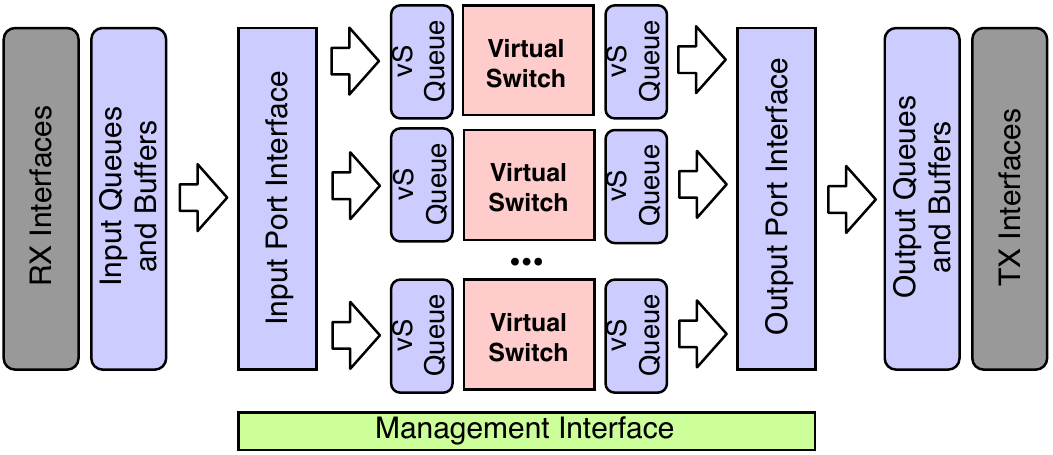}
    \caption{Conceptual architecture for programmable switch virtualization.}
    \label{fig:arch_overview}
    \end{figure}
    
    
    
    Our architecture was built on top of the canonical NetFPGA reference design~\cite{Ibanez:2019:PWL:3289602.3293924} and adapted for multiple switch instances according to P4VBox~\cite{p4vbox}. It virtualizes the forwarding plane, providing supporting structures for deployment and parallel execution of virtual switch (vS) instances. The switch code can be written, compiled, and ran independently to provide data isolation between switch instances and protect vendors' intellectual property. Fig.  \ref{fig:arch_overview} overviews the proposed architecture. It is composed of TX/RX network ports, an Input Port Interface (IPI), an array of vS instances (vS Array), an Output Port Interface (OPI), and a Management Interface (MI).
    
    
    The \textbf{vS Array} is a set of vS placeholders. It contains multiple preallocated circuit areas with a given set of resources, each capable of implementing a single vS instance. The vS instance is the core of our architecture. It actively performs packet switching. To enable vS packet forwarding, the vS placeholders contain three communication channels that receive packets from the IPI, send packets to the OPI, and process control requests from the external MI. They all have individual logic to implement a vS instance, with an individual set of networking protocols, switch metadata, variable scope, and control flow. They also have a private memory space for storing match-action tables and registers, which can only be accessed by the vS instance. The introduction of multiple vS instances creates the need to steer flows form the RX network ports to corresponding vS instances and then back to the TX network ports. The IPI and OPI modules are responsible for these tasks.
    
    The \textbf{IPI} buffers and serializes input packets to a parser. The parser searches the packet for a valid network segmentation tag to decide where to steer it. We currently use 802.1Q VLAN to this end. We then apply the packet to an abstraction of an \texttt{Ingress} table, which the manager configures through the control engine. It has two actions: \emph{forward} the packet to a vS instance or \emph{drop} it. The forward action matches the packet VLAN tag and receives a device id as an action field, indicating the identifier of the vS to receive the packet. vS instances may belong to the same VLAN, as long as they do not share the same physical ingress ports. In case the parser cannot find a valid device id, the packet is dropped.
    
    The \textbf{OPI} module delivers packets that have been processed by the vS array to a TX network port. It does so by iterating over vS instances' output buffers, forwarding packets as soon as the output TX port is available. It also applies the packet to an abstraction of an \texttt{Egress} table. As the IPI's ingress table, it also has two actions: \emph{forward} the packet to a TX port or \emph{drop} it. The forward action matches the packet VLAN and the \texttt{device\_id} informed by the IPI, to ensure that a vS does not forward packets to unauthorized network segments.
    
    
    The \textbf{MI} module provides a Command-Line Interface (CLI) between all architectural modules and the network administrator. It provides read/write access to all data structures, including tables and registers in the IPI and OPI and match and action tables on deployed vS instances. The control itself is not part of the platform implementation; instead, it lies on the software stack, external to the platform.
    

\section{Hybrid FPGA-ASIC Platform Implementation}
\label{sec_architecture}

    \begin{figure*}[ht!]
    \centering
    \includegraphics[width=1\linewidth]{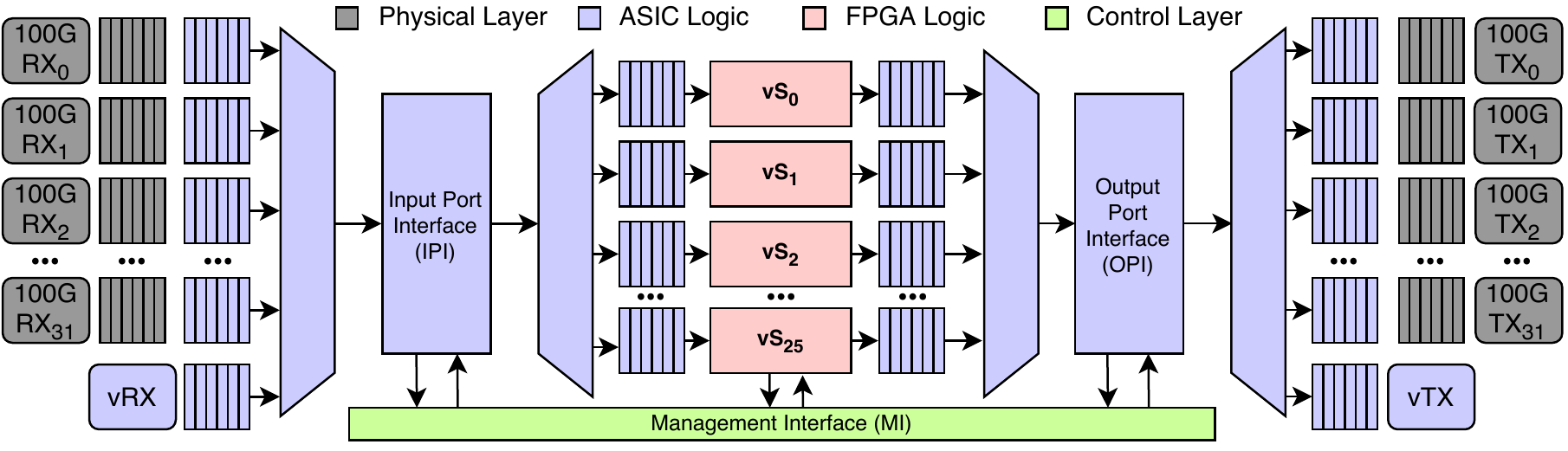}
    \caption{Overview of our hybrid FPGA-ASIC platform conceived as a System-on-Chip.}
    \label{fig:arch_implementation}
    \end{figure*}


We conceive our hybrid FPGA-ASIC platform as a System-on-Chip (SoC) that targets a 65 nm commercial CMOS process from a partner foundry. Fig. \ref{fig:arch_implementation} presents an overview of our platform. Red and purple blocks are part of the SoC (FPGA and ASIC logic, respectively). Grey blocks represent the physical layer, implemented using external ICs, providing the system with 32 100 Gbps RX/TX lanes. The green block represents an outer control layer, implemented off-chip, through a software stack running on an external microcontroller. A single input and 26 (one for each vS) output channels of 146 bits provide the middle ground between the MI and the chip.


To support reconfiguration, vS instances were allocated to an FPGA fabric, roughly equivalent to the Xilinx XCVU13P FPGA, while the MI was allocated to an external control layer. At full capacity, this device can support a vS Array with up to 26 virtual instances of the L2-switch, a popular case-study switch in the literature~\cite{Ibanez:2019:PWL:3289602.3293924}. Partial and full reconfiguration can be used to repurpose the switches' network function, and thus, the function of the system as a whole.


The IPI and OPI modules and their input and output queues and multiplexers are implemented using standard cells and memory IPs in the ASIC portion of the SoC. This approach keeps the implementation of memory-hungry portions out of the FPGA, thus allowing us to pick a reasonable, cost-effective IP. A total of 52 queues are needed to interface IPI and OPI with up to 26 virtual switches. Additional 64 queues are needed to interface the former blocks with the 32 RX/TX lanes. We also implemented one virtual channel through 2 extra vRX and vTX ports for loopback switching capabilities, resulting in 118 FIFOs being implemented. Our system partitioning approach establishes a trade-off where IPs are moved on the system according to IO count and bandwidth requirements~\cite{splitchip}.

    
\subsection{ASIC Logic Implementation}


    For the ASIC portion of the proposed system, we have developed specific verilog codes that describe, in RTL, the following blocks: IPI, MI, OPI, and the interfaces to drive/read RX and TX ports. We achieved the best performance at power trade-off by combined use of transistors of low, standard, and high threshold voltages (LVT, SVT, and HVT, respectively). Synopsys Design Compiler was our tool of choice for logic synthesis, targeting a 1GHz clock frequency, a demanding frequency for the 65nm node.
    
    A high degree of attention is given to the 118 FIFOs present in the design. Whenever possible, compiled SRAM blocks were used for packet storage in the queues. Our partner foundry also provides a SRAM compiler, describing it as a high-performance low-leakage single port compiler. Unfortunately, compiled SRAM comes in specific sizes/ratios that we have to adhere to. Our FIFOs have odd sizes: 52 x 289 (52 addresses with 289 bits) for the vS Array -- IPI/OPI and 66 x 417 for TX/RX -- OPI/IPI. The first example was implemented with flip-flops because the ratio is too asymmetrical: the compiler would not generate anything less than 32 addresses for the given word size. The second example is also non-trivial: 417 bits is not a power of two or even a multiple of two, for which we generated three instances of 64 x 144 to minimize the number of unused resources. 
    
    A total of 284 memory instances is required to provide the queues enough depth to achieve terabit bandwidth. The total number of SRAM bits is 2.54M (approximately 317K bytes). For all FIFOs but one, the storage is implemented with SRAM. All FIFOs have logic for write/read pointers and full/empty flags, which are implemented with conventional standard cells and operate as a wrapper around the SRAM memory instances. All FIFOs operate in asynchronous mode and are always on: this design choice increases area and power, but it is essential to enable the terabit bandwidth we seek.

\subsection{FPGA Logic Implementation}
    
    \begin{figure}[t]
    \centering
    \includegraphics[width=1\linewidth]{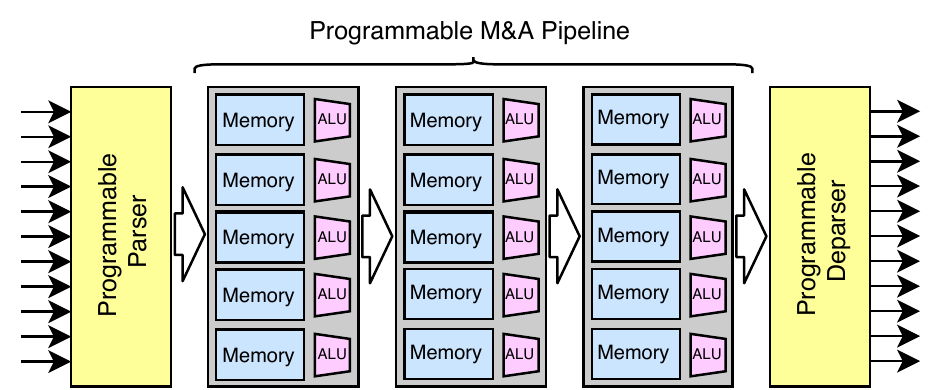}
    \caption{vS instance implementation}
    \label{fig:vs_implementation}
    \end{figure}
    
    The FPGA logic implements multiple vS instances in parallel and interface them with the ASIC logic. Fig. \ref{fig:vs_implementation} shows a vS implementation. A vS is generated through a High-Level Synthesis (HLS) method based on a model similar to the Very Simple Switch (VSS) reference model\footnote{https://p4.org/p4-spec/docs/P4-16-v1.1.0-spec.html\#sec-vss-arch} described in \cite{Ibanez:2019:PWL:3289602.3293924}. Each vS contains an ingress parser (Programmable Parser), multiple match-action pipeline stages composed of memory and ALU elements, and an egress parser (Programmable Deparser). Note that this is a suggested vS implementation flow; one could also write its own vS from scratch.


    To design and deploy a vS into the FPGA logic, one must describe a vS RTL that is able to communicate through one of the available vS queues in the ASIC logic and describe its packet processing pipeline. Thus, as long as designers wrap their packet processing pipelines, they can generate virtual switches ranging from a single wire to a full-stack router.
    
    
    To speed up the process of generating complex virtual switches, we employ an HLS design flow methodology to synthesize Domain-Specific Languages (DSL) such as P4~\cite{Bosshart:2014:P4}, Lyra~\cite{lyra-sigcomm2020}, and POF~\cite{Song:2013:PFU:2491185.2491190}. For generating a switch implementation from a P4 description, we take advantage of the  P4-NetFPGA project\footnote{https://github.com/NetFPGA/P4-NetFPGA-public}, which uses the Xilinx SDNet tool and the Simple Sume Switch model to automatically generate complex vS HDL. As the HLS tool is only compatible with Xilinx platforms, we emulated our FPGA fabric by targetting the Xilinx XCVU13P device.
    
    The vS implementation flow consists of wrapping a generic HDL-described switch with data and (optionally) control communication channels. To do so, the vS wrapper parses the HDL-described switch, inserts wrapper interfaces to communicate with the ASIC logic, and creates a final HDL file that can then be synthesized/translated into a target board configuration file. Additionally, the implementation flow can generate management drivers to access information about the switch pipeline, including the match-action tables, registers, and methods for accessing these data through the external MI.



    It is important that the HLS design takes into account the flexibility brought by network programmability, and make sure that network operators can redefine the forwarding plane behavior at will. Thus, the design must not impose static switch deployment. Instead, it must ensure network programmability's flexibility while maintaining high throughput. A reconfigurable FPGA is ideal for achieving both constraints, as it provides both flexibility and low latency. FPGA reconfiguration can be achieved by employing full or partial reconfiguration~\cite{FPGA}.


    With \emph{full reconfiguration}, the entire FPGA fabric is reprogrammed using a single bitstream that contains each HDL-described switch that must be deployed. This method reduces the need for resources to place vS instances, as the bitstream synthesis process can be optimized for the specific set of vS instances that must be deployed. It means, for example, better area occupation, because of an optimal placement of each vS in the fabric. As a result, full reconfiguration enables placing more vS instances. It also delivers higher operation frequencies for the entire fabric, meaning that instances will operate with higher throughput and lower latency.
    

    One problem with full reconfiguration is that  instances cannot be replaced (i.e., undeploy a  and deploy another one) unless the entire board is reconfigured. In this case, any running  instances in the board are terminated, and the context and buffers/queues of each  are lost. To solve this problem, \emph{partial reconfiguration} can be used. In summary, it involves partitioning the fabric into static and reconfigurable areas. Each reconfigurable area in the design can then be reprogrammed without affecting the operation of other reconfigurable areas. Each reconfigurable area can then be used for deployment of a  instance. The partial reconfiguration also enables faster deployment of a single instance, as only a portion of the fabric is reconfigured (in contrast to full reconfiguration).
    
    In spite of the flexibility of independent (un)deployment of  instances provided by partial reconfiguration, it requires more resources in the fabric for partitioning, therefore decreasing the number of vS instances that can be placed in the board. As the partitioning is done before vS deployment, each vS is constrained by the size of predefined partitions. More importantly, partial reconfiguration may also degrade the performance of vS instances, as the partitioning increases the distance between modules. It therefore brings critical implications to the deployment of switches that should achieve an aggregated throughput in the terabit mark.

\section{Results}
\label{sec_results}

    In Table~\ref{tab:asic_synthesis}, we show the results for the ASIC logic of our platform. All results are reported for a frequency of 1GHz, nominal VDD of 1.2V, and a TT (typical-typical) process corner. It is clear from the results that the memories dictate the area and power characteristics of the system. The memories represent 93.6\%, 90.4\%, and 92.8\% of the area, dynamic power, and leakage power budgets, respectively. Furthermore, out of the 718K cells (343K flops), approximately 42K are buffers, indicating that our targeted frequency is not trivial to achieve. An analysis of the critical path revealed that SVT cells are employed with modest driving strengths, mostly X1 and X2. Our results also revealed that the frequency could be pushed up to 1.3 $GHz$ for the TT corner, but at a high cost in both dynamic and leakage power since LVT cells (i.e., faster but leaky) would tend to be used more often.

    \begin{table}[hb]
\renewcommand{\arraystretch}{1.2}
\caption{Results from ASIC synthesis on 65nm technology.}
\label{tab:asic_synthesis}
\centering
\begin{tabular}{lrrrrr}
    \toprule
    F=1GHz, VDD=1.2V            & Logic     & Memory    & Total     \\
\midrule
Area ($mm^2$)                   & 3.0       & 44.6      & 47.6      \\
\# cells                        & 718227    & 284       & 718511    \\
Dynamic power ($W$)             & 2.7       & 25.6      & 28.3      \\
Leakage power ($\mu W$)         &  54.5     & 712.6     & 767.1     \\
\bottomrule
\end{tabular}%
\end{table}

    Table~\ref{tab:fpga_synthesis} presents data for the FPGA logic synthesis targeting the Xilinx XCVU13P FPGA. It shows the required LUTs, FFs, and BRAMs for each case-study switch implementation as well as achieved throughput for the minimum clock period of 1.392 $ns$ (718.4 $MHz$ clock frequency). The maximum clock frequency is directly related to the vS instances' buffers, implemented with BRAMs, where the critical path resides. Improvements to these structures could increase operating frequency, therefore improving overall throughput.


    In terms of area occupation, L2-Switch has minimum requirements, followed by Firewall, Router, and INT, which requires the most resources. Considering a XCVU13P, we could fit 26 Switches-l2, 14 Routers, 17 Firewalls, or 11 INTs in a single board. For these switches, the bottleneck is BRAM availability. They require, on average, double the percentage of available BRAMs as LUTs and FFs. Implementing BRAMs as ASIC could increase design efficiency and throughput.

    When considering latency and throughput, latency takes into account the time for a packet to cross the FPGA logic. In contrast, throughput considers the maximum amount of data passing through the FPGA logic. Thus, latency depends on the complexity of the vS instance's parser and deparser and the length of its match and action pipeline. The throughput, on the other hand, depends on vS latency and resource occupation. As an example, the maximum amount of L2-Switch instances running in parallel could achieve a maximum of 26 x 132.63 $Gbps$, or 3.448 $Tbps$. Note that a single L2-Switch instance can only achieve a maximum of 132.63 $Gbps$.
    
    \begin{table}[h]
\renewcommand{\arraystretch}{1.2}
\caption{Results from FPGA synthesis on XCVU13P.}
\label{tab:fpga_synthesis}
\centering
\begin{tabular}{lrrrrr}
    \toprule
    vS          & \multicolumn{1}{c}{LUTs}       & \multicolumn{1}{c}{FFs}        & \multicolumn{1}{c}{BRAMs} & \multicolumn{1}{c}{Throughput ($Gbps$)}    \\
    \midrule
    L2-Switch       & 27,626    & 39,520    & 102       & 132.63                    \\
    Firewall        & 48,979    & 76,147    & 153       & 130.92                    \\
    Router          & 49,754    & 80,915    & 185       & 131.45                    \\
    INT             & 77,956    & 155,594    & 240      & 129.61                    \\
    \bottomrule
\end{tabular}%
\end{table}



    Table \ref{tab:chip_throughput} presents the final results for the achieved throughput by our proposed platform. The platform's bottleneck is the Physical layer, with 32 100 $Gbps$ PHYs adding up to a 3.2 $Tbps$ throughput and, depending on the vS instances' complexity, the FPGA logic, which is constrained by its internal BRAM availability and varies between 1.43 and 3.45 $Tbps$. Note that additional PHYs could be added to the platform, and a different FPGA part could be used to increase maximum throughput. Considering 26 L2-Switch instances running on our proposed platform, we saturated the 32 100 $Gbps$ PHYs at 3.2 $Tbps$ in a single data direction, thus 6.4 $Tbps$ could be achieved in full-duplex mode.
    
    
    A comparison with related works is difficult because, to the best of our knowledge, this is the first platform in the literature to provide switch virtualization at the terabit mark. While the literature is still focusing on implementing virtualization on FPGA platforms~\cite{Ibanez:2019:PWL:3289602.3293924,p4vbox,Noa} at the gigabit mark, commercial approaches do not provide virtualization~\cite{TOMAHAWK,TOFINO,SILICON_ONE}. Therefore, our proposed proof-of-concept platform provides a combined virtualized throughput of up to 3.2 Tbps. This value is an order of magnitude higher than the available FPGA solutions but lower than the highest ranking commercial ASIC solutions at 25.6 Tbps~\cite{TOMAHAWK}. Still, one could further expand our platform's throughput by increasing the clock frequency on the ASIC-side and the configurable resources on the FPGA-side, at costs in area and power.

    \begin{table}[h]
\renewcommand{\arraystretch}{1.2}
\caption{Achieved Throughput Metrics for our SoC.}
\label{tab:chip_throughput}
\centering
\begin{tabular}{llrrrr}
    \toprule
    Logic          & \multicolumn{1}{c}{Instance}       & \multicolumn{1}{c}{\#}   & \multicolumn{1}{c}{Max. Throughput ($Tbps$)}    \\
    \midrule
    Physical        & 100 $Gbps$ PHY        & 32        & 3.20       \\
    ASIC            & AXI Stream            & 33        & 3.30       \\
    FPGA            & vS Instance           & 11--26    & 1.43--3.45 \\
    Chip            & Platform              & 1         & 3.20       \\
    \bottomrule
\end{tabular}%
\end{table}

    
\section{Final Considerations}
\label{sec_conclusion}


The dawn of 5G and the emergence of next generation networks will pose a major challenge to virtualization solutions powering public and private clouds. The prospect involves supporting slices of virtual networks capable of handling an aggregated traffic at the terabit scale and whose behavior can be redefined through forwarding plane programmability. However, programmable forwarding plane virtualization is still at is infancy \cite{p4visor,virtp4_sigcomm19,p4vbox,Noa}, with researchers still leveraging FPGAs to address aspects like resource sharing and isolation between slices of virtual programmable devices.

In this paper, we discussed the design of a hybrid FPGA-ASIC platform for switch virtualization, conceived as a System-on-Chip. In contrast to existing solutions, we leveraged increased clock frequencies with an ASIC, and combined it with the advantages of FPGA reconfigurability. Our experiments provide evidence of the potentialities of our hybrid platform, which is capable of achieving a combined virtualized throughput of 3.2 Tbps with low resource occupation overhead. Even though a comparison to related works is difficult, our results surpassed FPGA implementations by an order of magnitude while falling short to compete with the best available commercial ASIC solution, that does not provide switch virtualization, in terms of throughput.

In spite of the progresses achieved, much work remains. One direction for further research is ensuring that the pipeline of virtual switches can enable multiple access to counters, registers, and external functions, without degrading virtual switch throughput. We also intend to investigate how to leverage partial reconfiguration to enable the deployment of virtual switches with diverse memory occupation requirements.

\ifCLASSOPTIONcaptionsoff
  \newpage
\fi

\bibliographystyle{IEEEtran}
\bibliography{ref}

\end{document}